\documentclass[screen, nonacm]{acmart}

\usepackage[dvipsnames]{xcolor}
\usepackage{soul}
\sethlcolor{Apricot}
\usepackage{booktabs}
\usepackage{booktabs}
\usepackage{graphicx}
\usepackage{longtable}
\usepackage{makecell}
\usepackage{microtype}
\usepackage{hyperref}
\usepackage[type={CC}, modifier={by}, version={4.0}, imagewidth=0pt]{doclicense}

\AtBeginDocument{%
  }

\setcopyright{none}

\usepackage{enumitem}
\begin{document}

\title[Human Oversight of Agentic Systems in Practice]{Human oversight of agentic systems in practice: Examining the oversight work, challenges, and heuristics of developers using software agents}

\author{Shipi Dhanorkar}
 \authornote{Both authors contributed equally to this research.}
 \email{shipidhanorkar@gmail.com}
 \affiliation{
 \institution{Microsoft}
 \city{Redmond}
 \country{USA}
 }
\author{Samir Passi}
 \authornotemark[1]
 \email{v-sapassi@microsoft.com}
\affiliation{%
   \institution{Microsoft}
   \city{Redmond}
   \country{USA}
}

\author{Mihaela Vorvoreanu}
 \affiliation{%
   \institution{Microsoft}
  \city{Redmond}
   \country{USA}
   }
 \email{mihaela.vorvoreanu@microsoft.com}

\renewcommand{\shortauthors}{Dhanorkar, Passi \& Vorvoreanu}

\begin{abstract}
    Autonomous software agents hold promise to increase developer productivity but make mistakes and exhibit novel failure modes, making human oversight central to successful human-agent collaboration.
    Existing research on agent oversight is largely conceptual; normative frameworks exist, but how users \textit{actually} oversee agents is less known.
    In this paper, we bridge this gap by providing early empirical anchors for the theoretical discourse on agent oversight.
    Drawing on interviews with 17 experienced developers, we conduct an exploratory inquiry examining \textit{what} forms of emergent oversight work developers perform, \textit{when}, and \textit{how}.
    We also document the oversight \textit{challenges} developers face and the \textit{strategies} they have started using to address them. 
    We found at least four forms of emergent oversight work: a priori control, co-planning, real-time monitoring, and post hoc review.
    We show that oversight work is not only reactive and retrospective, as portrayed in existing research, but also preventative and proactive.
    We describe situated oversight challenges (e.g., difficulty reviewing agent-generated code) and outline heuristics developers adopt to address such challenges (e.g., using test results as guarantees for code correctness).
    We conclude with high-level takeaways, future research directions, implications for the human-centered design of software agents and for software engineering practice, and limitations of our research.
\end{abstract}

\settopmatter{printacmref=false}

\maketitle

\section{Introduction}
Agentic AI systems are becoming ubiquitous, with applications across tasks such as data analysis, scientific discovery, and web navigation \cite{agentsindatascience,gaicoscientist,scientificresearch,park2023generative, nakano2021webgpt, shen2023hugginggpt}.
Unlike conventional AI, agents can autonomously plan, use external tools, and take actions in real-world environments \cite{schick2023toolformer, agent-architectures, reactagentarchitecture, shinn2023reflexion}.
The impact of agents is perhaps most evident in software engineering, which has seen a proliferation of tools such as Cursor, Claude Code, Cline, ChatGPT Codex, and GitHub Copilot, and in technical advancements such as Open Code Interpreter and AlphaEvolve \cite{zheng-etal-2024-opencodeinterpreter, Alphaevolvecodingagent, Agents-survey2025wang}. Specialized \textit{code agents}--systems capable of making repository-level changes \cite{swe-agent2024yang, Codeagent, OpenDevin}--have pushed way beyond simple code generation tools \cite{basiccodegenerations, basiccodegenerations2} and made their way into developer workflows and organizational practices.

Agentic systems, however, are not without problems.
They often fail for several reasons, such as misinterpreting requests, errors during interactions with external tools, inability to identify mistakes, and context loss \cite{cemri2025multiagentllmsystemsfail, microsoft-agentic-failure, challenges-agent-communication-2024, salesforce-failures-2025}.
Autonomous coding accelerates software development but also paves the way for mistakes that silently propagate into production.
Irreversible errors, ranging from localized file deletions \cite{Userincidents-gupta} to large-scale production incidents, such as the Replit agent's database erasures  \cite{Replitincident}, pose a significant threat to system reliability.
Agents also face cybersecurity risks both as vulnerable systems (e.g., susceptibility to prompt injection attacks, as in Gemini CLI  \cite{googleagentexposesuserfiles}) and as enablers of large‑scale autonomous attacks (e.g., credentials and sensitive data leakage, such as in the Claude Code incident \cite{Claudesecurityincident}).
While automated safety methods (e.g., red teaming, agents‑as-a‑judge) are increasingly used to catch agentic failures \cite{feffer-red-teaming,zhugeagentasjudge,self-verification-limits}, they remain limited by the scenarios and incentives they encode \cite{raji2021benchmarklimitations}. 
Agents may exhibit biases \cite{Magenticmarketplace} and behave differently in real deployments than in evaluations \cite{needham2025evaluationawareness, METR-evaluationawareness}.

All of this makes \textit{human oversight} of agents a priority.
This is also evident in governance frameworks that increasingly mandate human oversight of AI.
The EU AI Act explicitly requires that humans retain meaningful control over high-risk AI, and the European Commission's guidance on trustworthy AI identifies oversight as a core requirement \cite{Oversight-EU-ACT}.
However, such frameworks assume that oversight is feasible without thoroughly examining or specifying what it looks like in practice.
It often falls on users to \textit{do} oversight in their context, ensure agents work properly, and fix agent mistakes and failures \cite{openai2023practices, Oversight-EU-ACT, Biden-2023-AI, koulu2020_humancontroloversight, cheruvu2025_inonloop}.
However, users frequently struggle to identify mistakes, even in conventional AI outputs, and exhibit overreliance \cite{review-burden-llm-2025, Passi-Appropriate-2024, Passi-AR-Book-Chapter-2025, passi2022overreliance}.
Agentic systems exacerbate these issues because they introduce new failure modes and require users to assess not only agent outputs but also agent actions \cite{openai2023practices, passi2025_agenticoversightproblem, microsoft-agentic-failure, cemri2025multiagentllmsystemsfail}.
Oversight challenges are particularly acute for developers who must
oversee agents across multiple software tasks and projects in production environments under time pressure.

Agent oversight is essential for effective human-agent collaboration and is increasingly seen as an accountability and governance requirement.
However, it is remarkably difficult to facilitate or even know \textit{how} to enable effective agent oversight \textit{in practice} \cite{challenges-agent-communication-2024, eric-aviation-2024, sterz2024_interdisciplinaryhumanoversight, green2022_oversightisflawed, Oversight-Possible-2025, passi2025_agenticoversightproblem}.
As Section \ref{sec: litreview} explains, most research on oversight is conceptual or narrowly focused on specific tasks, such as finding mistakes in AI outputs.
Even with the pressing need to examine on-the-ground practices, there is little empirical research that holistically examines how users \textit{actually} oversee AI or agentic systems in messy real-world settings, rendering current oversight efforts largely aspirational \cite{lai2023_scienceofhumanaidecisionmaking, passi2025_agenticoversightproblem, turri2024_transparencyinthewild, wang2024_trustaicode}.
We take qualitative, human-centered steps in this direction.
We focus on the agent oversight practices of software engineers because they are the professionals most likely to understand the limitations of the technology and the need for oversight.
We specifically focus on power users: early adopters in well-resourced organizations and heavy users who can signal emergent practices and challenges and help model oversight best practices.
Challenges they encounter are likely to be exacerbated for other user groups.
Drawing on research in FAccT and adjacent fields on human oversight and human-AI collaboration and the emergent use and challenges of software agents, we pursue two research questions:
\begin{itemize}
    \item How do developers oversee software agents in practice?
    \item What oversight challenges do developers face and how do they address them?
\end{itemize}
Our goal is to take exploratory steps to highlight the situated work of overseeing software agents. Our participant pool spans multiple organizations but is largely situated within a single tech company. Therefore, we acknowledge that we do not provide fully generalizable findings or an exhaustive taxonomy of oversight work. Our motivation is to provide early empirical anchors to bridge the gaps between current aspirations and the emerging practice of agent oversight. In the rapidly accelerating context of agentic software engineering, our granular and focused insights provide a crucial baseline for understanding the practical aspects of emerging agent oversight work.
\textbf{We make three key contributions:}

\begin{enumerate}

    \item We provide one of the first \textit{empirical accounts} of the lived reality of developers overseeing software agents. This is a much-needed step towards bridging the theory and practice of agent oversight.

    \item We present an \textit{exploratory} \textit{qualitative analysis} of the emergent forms of oversight work developers do in practice, \textit{when}, and \textit{how}, including its \textit{challenges} and the \textit{heuristics} they use to address them.
    
    \begin{enumerate}
        \item We find developers doing atleast four forms of oversight work—a priori control, co-planning, real-time monitoring, and post hoc review—using specific practices to achieve varied goals.
        We show that oversight is not only reactive (responding during execution) and evaluative (reviewing post-execution), as portrayed in prior research, but also preventative (configuring before prompting) and proactive (co-planning).
        
        \item We highlight oversight challenges developers face (e.g., difficulty reviewing agent-generated code) and document heuristics they adopt to address them (e.g., using test results as guarantees for code correctness).
        We find that developers opt for efficient, not perfect, oversight, surfacing disconnects between research aspirations for ideal human supervision and the practical realities of using agents.
    \end{enumerate}
   
    \item We outline future sociotechnical research directions, provide design implications to address developers' immediate oversight challenges, and highlight broader implications for the software engineering practice.
\end{enumerate}
The rest of the paper is organized as follows. Section \ref{sec: litreview} provides an overview of relevant research to contextualize our research focus, findings, and takeaways.
Section \ref{sec:methods} describes our data collection and analysis methods, including recruitment and participant information.
Section \ref{sec:findings} presents the empirical findings. Section \ref{sec:discussion} discusses high-level takeaways, future research areas, and design and practice implications.
We end with limitations of our research and ways to address them.

\section{Literature Review}
\label{sec: litreview}
\subsection{Human oversight of AI}
\label{sec:oversightofAI}
Although we focus on human oversight of novel software agents, our conceptualization of oversight draws from research on human oversight of conventional AI systems such as predictive algorithms and recommender systems \cite{Oversight-Control-2020, Oversight-Possible-2025, Oversight-What-When-2023, koulu2020_humancontroloversight, cranor2008_humanintheloop, filippo2018_meaningfulhumancontrol, cavalcante2023_meaningfulhumancontrol}.
While a detailed overview of that research is outside the scope of this paper (we do engage with FAccT research on AI oversight in subsection \ref{sec:facct_research}), two insights are helpful to contextualize this paper's findings and arguments.

The first insight is that reviewing AI outputs is considered key to overseeing any AI system.
Researchers debate the specifics of what constitutes--or \textit{should} constitute--oversight, but agree that spotting and correcting mistakes is a crucial part of it \cite{spotting-mistake-signal-2024, spotting-mistakes-algo-2023, review-burden-llm-2025, openai2023practices, passi2025_agenticoversightproblem}.
The second insight is that the field of human oversight of AI has significant empirical gaps.
Most discourse on oversight of AI, and now agentic AI, either has a \textit{narrow empirical focus} on controlled studies of mostly artificial verification tasks using proxy measures (unsurprising, given the  importance attributed to the oversight work of reviewing AI outputs) \cite[e.g.:][]{schemmer2023_explanationsreliance, goyal2023_backgroundexplanations, green2019_limitsofalgodecisionmaking, buccinca2021_cff, klingbeil2024_overrelianceexperiment, lee2023_counterfactualreliance} or is \textit{largely conceptual,} focusing on high-level arguments, frameworks, and theories of oversight \cite[e.g.: ][]{frenette2023_ensuringoversight, kyriakou2023_discoverai, eric-aviation-2024, cranor2008_humanintheloop, spotting-mistake-signal-2024, sterz2024_interdisciplinaryhumanoversight, Oversight-Control-2020, tssamados2025_supervisiontoteaming}.

The overall result is a disconnect between the research and practice of human oversight.
Empirical research is only beginning to explore how users \textit{actually} oversee AI systems in real-world settings.
Without this knowledge, it is difficult to identify \textit{on-the-ground} oversight work, practices, and challenges, let alone build \textit{effective} mechanisms to support oversight in practice.
We take steps to bridge this gap by providing exploratory qualitative analyses of the emergent oversight work, challenges, and heuristics of developers using software agents.

\subsection{Agents in software development: Applications and challenges}
The 2021 breakthrough demonstrating the ability of large language models (LLMs) to analyze and generate software code \cite{chen2021evaluatinglargelanguagemodels} sparked widespread interest, driving significant technical advancements \cite{shinn2023reflexion} including agentic frameworks such as OpenHands CodeAct \cite{wang2024executable}.
Coding agents have since been applied to a wide range of software engineering tasks \cite{hou2024litreview, lyu2025automaticprogramming}, such as debugging and patching \cite{zhang2024autocoderover,bouzenia2025repairagent}, code translation \cite{wu2025translation,eniser2024translatingtorust}, code refactoring \cite{horikawa2025agenticrefactoring,tornhill2025ace,CodeRefactoring-muarf2025xu}, and generation tests for bug reproduction or proof of vulnerability \cite{mundler2024testers,cheng2025bugreproductiontest,nitin2025faultline,ahmed2025otter2}.
Alongside technical advancements, developer productivity studies have demonstrated the positive impact of software agents on task time \cite{paradis2025googlerct,peng2023developerproductivity}. 
Sarkar’s \cite{sarkar2025productivity} natural experiment analyses measured another productivity metric, repository merge rates, and found that organizations using Cursor's agent achieved 26\% more merges than those that did not use it.
Coding agents continue to show impressive performance on benchmark evaluations, most prominently on SWE-bench and its related leaderboard \cite{jimenez2024swebench}.

However, software agents also exhibit critical limitations.
Studies have revealed severe flaws in LLM-generated code, including hallucinations \cite{Hallucinationssurvey-Huang,codehallucinations-AAAI,Packagehallucinations-Spracklen}, security vulnerabilities \cite{asleepathekeyboard-Pearce}, reasoning-action misalignment \cite{bouzenia2025understandingsoftwareengineeringagents}, and novel attacks originating from the MCP infrastructure \cite{MCPattackvectors2025song}.
Agents also falter on execution errors, such as database integrity violations and operating system errors that persist despite evaluations \cite{chen2025finalcodeprocessorientederror}.
The explosion of benchmarks such as SWE-Bench, RedCode, BigCodeBench, and CodeLM-Sec \cite{jimenez2024swebench,RedCodebenchmark2024guo,zhuo2025bigcodebench, CodeLMSec2024Hajipour} exposes coding agents' rapid evolution alongside their deepening flaws such as memory pitfalls \cite{liang2025swebenchillusionstateoftheartllms,memorization2024al-kaswan,memorization2024zhou}, test-passing but functionally broken patches \cite{solvedissues2025wang,zhu2025establishingbenchmarkbestpractices}, and failure to adhere to complex instructions with embedded constraints \cite{yan-etal-2025-codeinstructionfollowing}.
However, a gap remains: controlled evaluations do not reflect the high-stakes complexity of real-world software lifecycles.
This underscores the necessity of human supervision to mitigate risks and ensure real-world reliability in agent-driven software development.

Although formal research on developers' oversight practices is still lacking, we see glimpses of the need for and importance of human supervision in emerging human-centered research on agent-led software development.
A recent study made visible the benefits of developer revisions to agentic pull requests handling critical bugs, highlighting the need for human involvement to ensure the correctness of agent-generated code \cite{watanabe2025agenticcodingprs}.
Recent comparative studies between conventional AI coding tools and modern software agents document significant workflow shifts, noting that as agents take over task execution, humans tend to concentrate more on monitoring and debugging agent actions \cite{chen2025developerworkflowcomparisons,anthropicAnthropicEconomic}.
Kumar et al. \cite{kumar2025aiagentsneedyou} explicitly connect successful agentic outcomes with active, iterative developer involvement. 
Through field observations and a qualitative survey, Huang et al. \cite{huang2025notvibing} suggest that experienced developers rarely vibe code, relying on iteratively steering and editing agent outputs, because they regard one‑shot generations as unreliable.
Another research shows novice developers struggle to leverage agents effectively because of usability issues \cite{shome2025johnnycantuseagents}, highlighting tensions between industry visions of autonomous agents and developers' on-the-ground experiences and challenges.

\subsection{The emerging focus on human oversight of agentic systems in FAccT scholarship}
\label{sec:facct_research}
Human oversight of AI has also emerged as an important FAccT research topic, largely under the umbrella of human-AI collaboration, with researchers drawing connections between oversight and longstanding FAccT topics of \textit{accountability} \cite{donia2022_normativeaccountability, verhagen2025_moralagent}, \textit{auditing} \cite{casper2024_blackboxaudits}, \textit{fairness} \cite{henzinger2023_monitoringfairness}, \textit{harm} \cite{chan2023_agenticharms, weidinger2022_llmrisktaxonomy}, and \textit{responsibility} \cite{razzaq2024productivity, verhagen2025_moralagent}.
A common thread across this research is the link between oversight and transparency.
Meaningful oversight requires more than just placing humans in loops.
Users need ``epistemic access’’ to systems to adequately understand them \cite{sterz2024_interdisciplinaryhumanoversight}.
Transparency is key to enabling such access, and FAccT researchers continue to examine transparency's role and limits for human-AI collaboration (including the work of reviewing AI outputs) \cite{corbett2023_transparencyinfacct, henzinger2023_monitoringfairness, kaur2022_sensibleai, lima2022_explainableaccountable, sanneman2024_xaiinformationbottleneck, shang2022_counterfactuals}.
Shang et al. \cite{shang2022_counterfactuals} demonstrate how counterfactual explanations can improve human-AI decision making. Kim et al. \cite{kim2024_llmuncertainty} show how linguistic expressions of uncertainty in AI outputs help reduce overreliance.
Schuff et al. \cite{schuff2022_interpretingexplanations}, however, advise caution, arguing that users can misinterpret explanations because of biases and superficial factors.
Information provided by AI systems can be difficult to understand for some users and mislead others \cite{turri2024_transparencyinthewild}, exacerbating automation bias \cite{sterz2024_interdisciplinaryhumanoversight} and overreliance \cite{schmitt2024_roleofxaiforcollaboration, lee2024_multiplegenerationsoverreliance}. We draw on these insights when analyzing the information needs and practices of developers overseeing software agents. We contribute findings on the \textit{transparency challenges} and \textit{reviewing practices} of developers through a ``human-centered analysis [to further] define the design space'' \cite[p.1375]{lai2023_scienceofhumanaidecisionmaking} of human oversight and human-AI collaboration.

Much like wider research on human oversight, FAccT research has also mostly focused on conventional AI.
Researchers have only recently turned their attention to GenAI and agents, with early insights hinting that such systems exacerbate oversight needs and challenges.
Venkit et al. \cite{venkit2025_llmsearchissues} focus on the additional burden of verifying GenAI outputs. Wang et al. \cite{wang2024_trustaicode} show that GenAI systems often lack adequate support for output evaluation.
Agents introduce further risks; e.g., ``reward hacking'' makes agents produce right outputs for wrong reasons \cite{chan2023_agenticharms}, making it necessary for users to review both the outputs \textit{and} working of agentic systems \cite{chan2023_agenticharms, verhagen2025_moralagent, specification-gaming-2020}.

Addressing agentic oversight requirements and challenges, Chan et al. \cite{chan2024_visibilityintoagents} posit, requires designing new forms of ``visibility'' into the complex composition and working of agentic systems. Concurrently, such new designs must also account for the trade-offs  additional oversight aids introduce between system understanding, user workload, and human cognition \cite{sanneman2024_xaiinformationbottleneck, passi2025_agenticoversightproblem, naik2025_earlyagentadopters, openai2023practices, overthinking-2025}.
As researchers in FAccT and adjacent fields have recently pointed out, there is a pressing need for more empirical, human-centered research to examine how users navigate such tradeoffs when working with agentic systems in practice, likely ``compensat[ing] for unmet'' needs \cite{turri2024_transparencyinthewild} through oft-invisible situated actions and intuitions \cite{lai2023_scienceofhumanaidecisionmaking, chen2023_xaihumanintuition, wang2024_trustaicode, turri2024_transparencyinthewild, passi2025_agenticoversightproblem, epperson2025_debugsteer, brachman2025_agenticmentalmodel}.

Our exploratory analysis of agent oversight in practice is a step toward this goal.
We contribute to FAccT research on agent oversight and human-agent collaboration by analyzing and documenting the different oversight work developers perform, the challenges and tradeoffs they face when doing different oversight work, and the heuristics developers use to navigate tradeoffs and work around such challenges.

\section{Research Methodology}
In the subsections below, we provide details on our participation criteria, recruitment process, participants, data collection, and analysis strategy.

\label{sec:methods}
\subsection{Participants: Participation criteria, recruitment process, and organizational context}
We followed a criterion-sampling strategy based on experience using software agents for professional work.
In this study, we define a \textit{software agent} as one that autonomously plans and acts to accomplish user goals by using tools and interacting with the environment.\footnote{Given this definition, our work is informed by existing research but has a distinct goal.
For example, we extend Chen et al.'s \cite{chen2025developerworkflowcomparisons} research identifying oversight as a new form of work with the emerging mechanics of that work and Wang et al.'s \cite{wang2024_trustaicode} research on trust in modern AI systems with a focus on autonomous agents.}
Our recruitment materials included examples, such as Claude Code and Cline, to clarify what we meant.
We did not consider an AI that just auto-completes code as an agent.

We first reached out to a few professional contacts, and employed snowball sampling to identify additional participants.
Participants filled out a screener to determine participation eligibility.
We assessed our sampling criteria through a combination of participants’ self-reported \textit{agent experience} and \textit{agent usage for professional work}.
We only selected those who were \textit{somewhat experienced} with agents and used them at least \textit{weekly} for professional work.
Sixteen of 17 participants used agents several times a week for professional work (12 daily).
Nine participants reported being `experienced' with agents, four `somewhat experienced,' and four `very experienced.'\footnote{Since agentic AI is still an emerging tech in its nascency, we deliberately chose to not ask for participants' agent experience in months/years and instead opted for open-ended language in the screener (e.g., `somewhat' and `very' experienced).
To further ensure we get relevant participants, we used the interview’s opening questions as a secondary screener to assess participants' conceptual understanding and usage of agents.
We excluded one participant based on secondary screening.} 
See Appendix \ref{sec:appendix_participant_demograhics} for further participant details.

Ours is also an intensity sample \cite{patton2014qualitative}, concentrating on experienced developers who likely understand agentic AI better than people in other professions.
Thirteen participants had seven or more years of programming experience, two had over four years, and two had more than a year, spanning domains such as ad experimentation, financial technology, research, and storage systems.
Challenges seasoned developers experience are likely worse for practitioners with less accurate mental models of agentic AI.
We also specifically focused on power users in well-resourced organizations, as early adopters and heavy users are well-positioned to signal emergent practices and help model best practices.

Twelve of 17 participants work at the same large-scale tech organization as the authors.
This occurred unintentionally due to participant recruitment challenges.
We wanted to interview experienced software agent users, but a large part of that user base consists of developers working in the tech industry.
NDAs, proprietary nature of agent work, and the competitive agent landscape made developers from  other firms hesitant to participate in our study.
Software agent users at our organization felt more comfortable participating and sharing details.
Organizational culture impacts practices but norms can vary across teams in the same company.
The 12 participants were on distinct teams across domains with no overlapping projects; they shared similarities (e.g., security requirements) but also exhibited differences (e.g., using different agents for tasks).
The 5 participants from other organizations shared similar experiences.

\subsection{Data collection and analysis}
We used a semi-structured approach for interviews.
Rather than present a preconceived definition of oversight, we let participants explain what agent oversight meant to them and how they conducted it.
We first asked developers to describe what they used agents for (see Appendix \ref{sec:appendix_agent_usage_scenarios} for agent usage details).
We then dived into how developers worked with agents, evaluated agents' work (e.g., spotting mistakes in agent outputs), and the difficulties they faced in doing so.
The structure of our findings (i.e., four forms of oversight work) is shaped by our analytical lens (i.e., explicitly asking users what they do before, during, and after agent execution).
We deliberately used temporal anchoring to ground participant responses in end-to-end agent workflows.
The structure of our findings is thus best seen as a useful organizing device and not an exact representation. 
Note, however, that participants did organize their responses temporally even without prompting when differentiating between the oversight work of \textit{co-planning} and \textit{real-time monitoring.}
See Appendix \ref{sec:appendix_interview_topic_guide} for details on interview questions.

Interviews were conducted 1:1 by the first author via video conferencing software, video-recorded, and transcribed using Marvin  \cite{heymarvinMarvinAINative}.
Participants completed a consent form prior to the interview.
Interviews were scheduled for one hour, lasting 60 minutes on average, and conducted between July-August 2025.
Participants received a US \$40 gift card or equivalent for participation.
The research study was approved by our organization's institutional review board.

We followed a qualitative inductive analysis process guided by our research questions and in concert with existing literature \cite{tracy_chapter_2024}.
Analysis proceeded in three rounds.
In round one, the first author conducted open coding, creating a list of developer actions, practices, and challenges through in-vivo codes directly using participant quotes (e.g., ``set deny lists'' and ``co-plan with agents'').
In round two, the first author captured common themes across open codes, creating 13 axial codes (e.g., \textit{co-planning} and \textit{review strategies}).
In round three, the first and second authors conducted selective coding, developing the final two conceptual categories that form our findings: different forms of oversight work (subsection \ref {sec:oversightforms}) and the heuristics used to work around oversight challenges (subsection \ref{findings_heuristics}).
The first and second authors regularly met during the second and third rounds to compare and resolve ambiguities.\footnote{For example, initially the authors differed on whether to keep \textit{a priori control} (subsection \ref{sec:oversightforms_aprioricontrol}) and \textit{co-planning} (subsection \ref{sec:oversightforms_coplanning}) as distinct categories or merge them into one ``control” category. On revisiting the coded data, a functional boundary between the two emerged: configuring pre-established controls vs. actively steering agents. Consequently, these were maintained as distinct axial codes.
In another instance, the authors initially differed on the nature of heuristics — one saw them as risky shortcuts while the other saw them as coping mechanisms. This was resolved by revisiting coded data to analyze participants’ motivation in employing heuristics, revealing them as imperfect-yet-necessary shortcuts to do \textit{good enough} post hoc review.}

\section{Findings}
\label{sec:findings}
Our data analysis process yielded two key themes--one related to emergent forms and challenges of oversight (subsection \ref{sec:oversightforms}) and another to the situated practical strategies developers use to perform oversight work despite its challenging nature (subsection \ref{findings_heuristics}).
Appendix  \ref{sec:appendix_agent_usage_scenarios} describes participants' agent usage scenarios to help contextualize the emergent agent oversight work we describe in this section.

\subsection{Four forms of human oversight work}
\label{sec:oversightforms}
Our exploratory analysis revealed four forms of emergent oversight work that developers perform when using software agents: a priori control, co-planning work,  real-time monitoring, and post hoc review (see Appendix \ref{sec:appendix_forms_of_oversight_table}).

\subsubsection{\textbf{A priori control}}
\label{sec:oversightforms_aprioricontrol}
A priori control is a form of oversight work that involves giving instructions to agents to direct and limit their workings \textit{before} delegating tasks to them.
The goal of a priori control is to define clear boundaries for agents and ensure they are set up properly to minimize failure.\footnote{Despite a priori contol, developers \textit{``never know''} for certain how agents will behave, leading developers to resort to working in \textit{``very secure environments'}' such as a \textit{``sandbox''} (P16) to experiment and understand the agent's working.}
Developers exercise a priori control using built-in agent settings and custom instructions.
A priori control includes configuring agent autonomy, stopping agents from using specific code libraries, and supplying global context to agents on project scope.
\begin{quote}
    ``There’s one essential thing in this puzzle: your instruction. […] A specific file in the repo that contains base instructions to understand the repo for the agent. That’s the most important asset […] that you’re constantly maintaining, improving, and iterating.'' (P15)
\end{quote}

A priori control is especially important because of the inherent unpredictability of agents--\textit{``sometimes [the agent] follows your instructions, sometimes it tries to go on a tangent and do its own stuff''} (P11).
The specifics of \textit{what} developers do for a priori control depend on the agentic failures they imagine or have previously encountered.

All participants acknowledged the need to constrain agents but differed in the nature and extent of the a priori control work they performed.
Some \textit{``spen[t] a lot of time setting up or configuring''} agents (P17), specifying detailed \textit{``dependenc[ies] and external contracts''} (P14). P03 explained, \textit{``I will try to set it up as much as possible, [e.g.] I update the deny list where I don't want the agent to ever delete something.''}
Other developers simply relied on default agent controls:
\begin{quote}
    ``I have seen some videos where you can redefine some actions and flows, but I have really not gone too much into that. [...] I just use it. `Agentic mode' is the default mode. [...] I just use that. I have not messed around with the configurations too much, to be honest.'' (P13)
\end{quote}

\paragraph{{\textit{\textbf{Challenges in a priori control}}}}
We observed two challenges to effective a priori control.
The first is that despite the presence of a priori control mechanisms, developers perceive having little control over an agent's working.
As P04 said, they \textit{``don't think [they] have''} much control over an agent's \textit{actual} operation--a sentiment echoed by several participants.
One reason for this perception is that participants primarily used prompts, and not custom instructions, to control agents, even though experienced developers argued that using custom instructions \textit{``drives directly the quality of what the agent produce[s]''} (P15).
Using black-boxed agents made matters worse:

\begin{quote}
    {``I didn't set up the exact agents that we use to assign tasks.
    We had to request from another team to have it integrated. I don't think we choose which specific model. It's probably their team that set that up, and we just sort of get the end-user impact of having the agent working.''} (P02)
\end{quote}
The second challenge is that developers must make informed choices while working with limited information about agents.
For example, P16 said they were \textit{``not even sure what reasoning [agents] are using.''} Beyond agent operation, developers also have limited information about agents' data policies--e.g., what \textit{``happen[s] to all this data that is going [in]to these models''} (P01), \textit{``especially if it is […] customer-related data''} (P03) being transmitted to third-party platforms.
    
\subsubsection{\textbf{Co-planning work}}
\label{sec:oversightforms_coplanning}
Co-planning is a form of oversight work that involves preliminary goal-setting activities that agents and developers jointly perform to establish common ground.
The goal of co-planning is to ensure agents identify key actionable requirements for successful task completion, minimize misalignment with user intent, and facilitate downstream oversight.
Examples include iterative prompting, drafting plans with agents, and seeding partial solutions to agents.
Co-planning is important because letting agents operate completely hands-off often results in their failure to accurately fill decision gaps independently.
\begin{quote}
    ``I learned as a hard lesson: do not completely go hands-off. [...]
    LLMs are short-term memory thing. [...]
    No problem can be solved only in one way, right? It could be solved in thousands of ways, and it will start flipping between those thousand ways, creating a mess because decisions are not consistent across the project. Sometimes it will make a decision—an important technical decision—in one part, and hours later, it will just forget about it and start again and make some other decision. It will be a hodgepodge of a project. It will be very difficult to make it work.'' (P05)
\end{quote}
Participants described two key co-planning practices.
The first is drafting plans with agents.
Through this collaborative work, agents and developers align on how best to accomplish goals by, for instance, building a shared understanding of the solution's \textit{``data flow, structure, key components, [and] how they're linked''} (P07) .
The desire to minimize review and repair downstream motivates developers to invest time and effort up front during co-planning: \textit{``While we are iterating software with these AI coding agents, it's better to have a design up front compared to making code changes later''} (P14).
This is especially important when there are many possible solutions and developers want agents to implement specific ones: \textit{``if there are multiple approaches, I tell what is the correct approach, and I ask it to follow the [company's] coding practices''} (P08).
Developers often seed partial implementations during co-planning to further guide agents:
\begin{quote}
    ``Sometimes, in the planning mode, it's not getting to the point that I want to. Then I write the code snippet. I bite the bullet. I finally write the code and say, 'Okay, this is the code snippet.' I don't write full code. I'm still a lazy developer. [...] I write the crux of it.'' (P11) 
\end{quote}
The second co-planning practice participants described is task decomposition.
Developers \textit{``deconstruct''} tasks into small \textit{``chunks''} where each \textit {``has minimal side effects and can be testable independently''} (P17).
Smaller tasks help \textit{``contain''} (P05) and \textit{``narrow down''} (P14) code changes to fewer files and scoped functions, making review and change easier.
As P05 explained, with smaller tasks:
\begin{quote}
    ``the margin of error is very low...this function is expecting this outcome in this format and providing this parameter. I know what to do with this. If that gap is bigger and it has to take a decision and it doesn't have context of the rest of the application, it would more likely build something that is not well integrated with the rest of the application.'' (P05)
\end{quote}
Coherent mental models facilitate co-planning.
When decomposing tasks, knowing when minimum details can suffice (e.g., when there are \textit{``well-defined, state-of-the-art solutions''} (P06) agents can easily find) and when agents require detailed instructions (e.g.,  where there are \textit{``crazy risk conditions based on logs, metrics, [...and] you need to provide all this context to the AI''} (P16)) is key to breaking down tasks effectively to steer agents towards success.
Similarly, when drafting plans with agents, a \textit{``sense of familiarity [for] things that [the agent] gets wrong vs. things it gets right almost all the time''} (P13) helps make informed decisions about guiding agents.

Overall, participants called out the iterative nature of co-planning work that includes, for instance, not only providing details, but also reviewing and refining plans: \textit{``once you have received a response, if it was like 10\% better than the previous response, you have to go on to more specifics''} (P04). Co-planning involves negotiating with agents: \textit{```Is this right?,' `why do you think we need two classes?,' and `why do you lay out folders this way?'''} (P14).
Participants described co-planning as an emerging form of \textit{spec-driven development} (P05, P15) that ensures agents remain \textit{``grounded''} (P05, P07): 
\begin{quote}
    ``First you write specs, then you have your agent that explains [... them,] does deep research on your issue [...,] runs, and then you ask further questions to that agent until you've built sufficient context. Then you run a last agent that does a [final] summary and creates a prompt that you use to launch the software agent.'' (P15)
\end{quote}

\paragraph{\textbf{\textit{Challenges in co-planning}}}
We observed two key challenges to effective co-planning.
First, it is difficult to identify the appropriate level of specificity at which to instruct agents.
What is \textit{``obvious''}  to humans is often invisible to agents (P01).
Experienced developers do not just solve coding problems but also consider implicit requirements such as security, performance, and maintainability (P04, P06).
However, unlike humans, who can infer missing information and adjust on the fly, agents make brittle decisions without detailed instructions.
Under-specified plans cause agents to misinterpret goals and overlook constraints (P03, P05, P11).
However, developers \textit{``don`t always know in advance''} (P11) how much hand-holding agents need or at what point hand-holding work overshadows the benefits of using agents altogether (P02).
\begin{quote}``If you provide a high level plan, you get high level code. [...] You tell me that I'm going to go from X to Y, but then I'm going to stop at T for a break. You implement a solution like X, Y, T. You say stop at T, but the coding solution shouldn't be that. The coding solution should introduce a new parameter: how much time are you going to stop at T?'' (P04)
\end{quote}
The second challenge is that using natural language for prompting makes it  difficult to appropriately articulate and specify goals to agents, widening alignment gaps between developer intent and agent working.
For simple things, developers often find themselves \textit{``over explaining''} to agents: \textit{``I will say things like, `do this in Python,' even though to me it's obvious that it's Python''} (P01).
In complex situations, developers frequently second guess whether their intent is clear enough for agents to interpret or whether the provided examples are sufficient for agents to identify the correct solution.
\begin{quote}
    ``Making the prompt coherent is like writing a story.
    Constructing that story in my head in a coherent way is not easy.
    I can have bits and pieces of that story.
    Code should do this, do that.
    [But] having a flow of story that if this happened, do this, that's not an easy task for a human.
    [...] Articulating stuff in the time you write the prompt, like 15 seconds, [...] that's my challenge, I think a universal challenge.'' (P06)
\end{quote}

\subsubsection{\textbf{Real-time monitoring}}
Real-time monitoring is a form of oversight work that involves observing agents as they perform tasks and intervening when needed.
The goal of real-time monitoring is to ensure agents remain on track, are not stuck in loops or failure modes, and follow predefined plans. Examples of real-time monitoring work include observing action logs, reading reasoning traces, and pausing and stopping agents mid-execution.

Participants revealed that they rarely performed real-time monitoring, with most never \textit{``proactively look[ing] at the [agent's] thought process’’} (P12).
The few times participants monitored action logs or reasoning traces, they did so perfunctorily: \textit{``eyeball[ing] [them] to see that [the agent] did the right thing based on what it said''} (P13).
Especially for longer running and complex tasks, relying solely on quick scans is problematic, causing developers to miss important details and allow subtle agent errors to slip through.

We hypothesize that a key reason for our participants not engaging in real-time monitoring lies in how they assigned tasks to agents.
As described earlier, developers decomposed large tasks into smaller sub tasks during co-planning to improve the intelligibility of intermediate agent outputs and facilitate post hoc testing.
Delegating small tasks to agents meant that agents worked quickly on mostly simple tasks, making it relatively easy for participants to make do either with quick glances during execution or with skipping real-time monitoring altogether and relying only on reviewing final agent outputs.

\paragraph{\textbf{\textit{Challenges in real-time monitoring}}}
Since participants rarely monitored agents in real time, they did not describe specific challenges with it. However, our analysis revealed three observations about the \textit{general} difficulty of real-time monitoring as described by our participants.

The first observation is on the consequential disconnect between what an agent says it is doing and what it actually does.
For example, developers regularly deal with agents' poor confidence calibration in reasoning traces: \textit{``10\% of the time it is wrong [about why it did something]. It is very confident, so you don't get to know.''} (P10).
Therefore, while artifacts such as reasoning traces and Chain-of-thought (CoT) rationales provide monitoring capabilities, they remain inherently unreliable.

The second observation is on the use of ad hoc cues to identify possible issues during real-time monitoring.
Participants relied on indirect cues such as abnormally long agent run time or conversation turns, and not vigilant monitoring, to spot issues such as context drift during execution (P03, P06, P17).
\begin{quote}
    ``One of the biggest things I observed [...] is that if you go a certain distance with the agent, then its hallucination starts growing. Like it has too much context over a period. If you spend 30 minutes or one hour, it has a big context. It starts doing random things. Things I didn't even ask.'' (P03)
\end{quote} 
The third observation is on the difficulty of fixing agent mistakes during execution.
When developers spot issues mid-execution, they work to address them and bring the agent back on track by, for instance, pausing the agent and manually updating its context or prompting it to fix mistakes before moving on.
However, participants mentioned (P02, P17) that such approaches were inefficient and unreliable:
\begin{quote}
    ``I find that if you try to redirect [the agent] and give it the right context, it doesn't get back on track. I have to sort of stop it and  then update my prompt and restart all over again.'' (P02)
\end{quote}

\subsubsection{\textbf{Post hoc review}}
Post hoc review is a form of oversight work that involves reviewing and correcting agent outputs after execution.
It was the most discussed oversight work in our study.
The goal of post hoc review is to ensure agent outputs are correct and are achieved using appropriate approaches.
Examples of post hoc review work include verifying agent outputs, making adjustments when agents falter, and re-prompting agents to fix issues.
Post hoc review is important because agents can not only make legitimate mistakes (e.g., incomplete code that does not handle all edge cases), but also hallucinate (e.g., mentioning \textit{``classes that don’t exist in codebases''} (P05)).
Developers conduct post hoc review through different practices such as reviewing code outputs using diff tools, using LLMs-as-a-judge, and layered testing (e.g., agent-generated test suites or manual testing of code functionality). 
\begin{quote}
    ``After the code gen, I get into the code diff [...] and start going through each file. It's how I review others' code: get to each file and, as with code reviews, you first see overall things, then you figure out some key features or code areas that it has modified. Go to those first, see if it's good or not, then start updating a few smaller things.'' (P11)
\end{quote}
The extent of post hoc review work depends on the type of tasks developers delegate to agents.
For instance, consider the difference between incremental and ground-up software development.
Incremental software development is entrenched in legacy codebases and intertwined with interconnecting modules.
\textit{``There [is] a lot of legacy code, a lot of modules, so the complexity becomes more because it's not generic.''} (P04).
Using agents to integrate new features in existing codebases makes rigorous post hoc review work even more crucial, at times \textit{``more stringently than if it was a person [who wrote the code]''} (P02).
\begin{quote}
    ``I read every single line.
    I approve every single line.
    [...] For serious software engineering, especially where you're part of a broader system, it's extremely problematic if you have code where you're not exactly sure what every single line does. You show up in a meeting, there are people who need to make integration decisions that rely on you knowing exactly how this code works.'' (P17)
\end{quote}
Unlike incremental software, ground-up software has little dependency on existing codebases or prior modules.
Examples include proof-of-concept and research prototypes.
Despite valuing post hoc review for functional validation, participants admitted granting their agents \textit{``unrestricted''} (P16) autonomy and maintaining reduced scrutiny in such cases.
\begin{quote}
    ``I do a lot of prototyping.
    [If] I just want something that will let me demo but I don't really care about the design.
    In that case, I'm okay with letting the agent just run wild.'' (P01)
\end{quote}

\paragraph{{\textbf{\textit{Challenges in post-hoc review}}}}
We observed developers facing two key challenges that inhibit effective post hoc review.
The first challenge is that using agents makes developers cognitively distant from the code they must review.
This increases the \textit{``cost or effort [developers must] put into verification, testing, [and] benchmarking''} (P17).
As agents code, developers drift further from the practice of writing code.
\textit{``A computer [is] writing the code, but humans need to read it''} (P14).
However, as P01 put it, \textit{``because [the code] is not created by me, my understanding of it is very surface level.''} \textit{``It's not your code you're reviewing, it's somebody else's. I can review my code much faster. It's like similar stuff with your handwriting. It is much faster for you to read [your own] than somebody else's handwriting’’} (P11). The growing distance between developers and agent-generated code makes it harder to troubleshoot even simple issues in agent outputs, especially because the sheer ``volume of [agent-generated] code is so high'' (P13).
\begin{quote}
    ``A couple of times I wanted to make some changes. [...] A very small change, like font size or something. [...]  I ended up spending quite a bit of time understanding the code, and it was a lot harder to understand because I didn't write it.'' (P01)
\end{quote}
The second challenge is that developers must re-review agent-generated code with every iteration (P11, P14).
When developers identify an issue with agent output and ask the agent to fix it, they feel the need to review the whole code again because the agent might have made changes, and at times unnecessary changes, in multiple places. 
\textit{``Each time you say, `okay, generate again or fix it again,' you have to review it again''} (P11).

\subsection{Four heuristics developers use for post hoc review}
\label{findings_heuristics}
The previous subsection showed different oversight work and highlighted several challenges.
Our analysis revealed that faced with the complex reality of overseeing software agents, developers have begun adopting heuristics that prioritize efficiency over perfection to work \textit{around} the challenges.
Below we describe four such heuristics participants used to accomplish the oversight work of \textit{post hoc review}.
We focus only on post hoc review because it was the most discussed oversight work among participants, and thus heuristics used for it were most prominently visible during analysis.
\paragraph{\textbf{Heuristic \#1: An agent's plan is a faithful proxy for its actual working.}} 
Participants heavily relied on the plan during post hoc review, treating it as a faithful representation of the agent's actual working and equating the quality of the agent's output with the seeming quality of the agent's plan  (P05, P07, P14, P15, P16).
This heuristic assumes that agents adhere to plans during execution and that an agent's working is congruent with plan contents, enabling developers to substitute the large effort of code review with the easier task of plan verification.
We see this most saliently at play in P16’s argument against the very need to understand an agent's ``thought process'' during post hoc review:
\begin{quote}
    ``When I ask [an agent] to do something, I don’t need to understand the whole thought process. It’s there, which I love, right? It’s doing something so that I get a feel of its thinking. But I don’t read through every single thing it does. The only thing that matters is the end: what did you do? [...] I just look at the plan, and then if I don’t like something, I mention it. I don’t care what it does in between. Whether it’s thinking, searching, looking. I don’t go through it line by line. That’s what everyone does. Maybe there’s a different way of doing it, but that’s just how I do it.'' (P16)
\end{quote}
The fact that developers spend a lot of up front effort co-drafting plans with agents is further evidence of this proxy-based approach.
Developers are willing to even change their workflows, adopting spec-driven development approaches assuming that agents will adhere to plans:
\begin{quote}
    ``I have to watch, understand, critique every outcome or steps it is taking, which is not scalable if I have to move fast. [...] What I pivoted to is creating a system, an approach that will help me do that instead of me doing that manually. This spec-driven development is part of that system: hey, let's solidify PRD before getting to tech spec; solidify tech spec before getting into UX; then create a full implementation plan before we even write a single line of code; then create task sections and tasks that LLM can blindly follow and get the right output.'' (P05)
\end{quote}
\paragraph{\textbf{Heuristic \#2. Passing test results guarantee the correctness of agent-generated code.}}
Participants often outsourced verification to the test suite.
When agent-generated code passes testing, it lends confidence in the code's correctness--the agent \textit{``did the job''} (P03).
In contrast, when tests fail, developers treat them as a way to identify which parts of agent outputs are problematic and must be evaluated during post hoc review. As P11 mentioned, going through test results instead of inspecting every line of agent-generated code is highly efficient:
\begin{quote}
    ``[When] the rubber meets the road, files get updated and I review the code, run the test cases. All the time, what I do is, after [the agent] writes code, I always [...] and we have agreed on test cases... I tell [the agent] to also ensure that you run tests and stuff and show me test results. Because it's easier to review test results and see it's bad, not working, and then iterate on it [...] than for me to go through the code.'' (P11)
\end{quote}
As several participants mentioned, the viability of this heuristic--i.e., reviewing test results in place of actual code--depends on the quality of tests themselves (notably P08, P11, P14). Indeed, as P08 surmised, it is possible to fully eradicate the need to review \textit{any} agent-generated code if developers can guarantee good test cases:
\begin{quote}
    ``We can ask AI or a human to write test cases, but humans need to review and be 100\% responsible for test cases. We need to make sure that the spec 100\% captures what the PM or the designer really wants. If that can be guaranteed, we can treat the source folder as a complete black box. If AI is able to pass all test cases and, of course, build, then we give it a go and we don't even need to take a look at the source folder anymore. When there's new changes, new requirements, just update test cases and let AI regenerate everything. There's no attachment that we have to the code.'' (P08)
\end{quote}
\paragraph{\textbf{Heuristic \#3. Eyeballing agent-related information, including but not limited to agent outputs, can reliably signal issues.}}
A main way in which participants managed information overload during post hoc review is through spot-checking by eyeballing agent-related information such as agent outputs, agent rationales, and change summaries.
The assumption underlying this heuristic seems to be that eyeballing suffices as an incomplete-yet-efficient information processing mechanism during review.
As P07 described: \textit{``I have this way of doing spot checks or high-level checks to make sure [the agent] is not missing some important files in the analysis.''}
P13 made a similar observation:
\begin{quote}
``Before I push any code, I look at all method signatures and give it an eyeball: `Okay, is this method needed?' [If] I see a method that I don't think makes sense over here, I'll just ask, `Is it actually needed? Should we remove it?' [... The agent also] gives me the reasoning and then does some work, then gives me another piece of reasoning, does some work. I can see the work it is doing is directly related to the reasoning that it is giving me. At the end, it summarizes: ‘these are the changes I made; what is the logic behind it?’ I can easily eyeball it and see, okay, it did the right thing based on what it said.'' (P13)    
\end{quote}
We observed developers not just eyeballing existing information but also asking agents to generate \textit{new} information such as visual diagrams to guide and facilitate post hoc review.
\begin{quote}
    ``I will ask the agent to go through the entire code path and build a data flow diagram for me. Go through all function calls and follow data flow of the entire pull request and generate a Mermaid-compatible diagram. This will give me the pseudo code so I can copy and put it in the Mermaid visualization tool to view the structure and data flow. This is really helpful for big PRs to understand whether all the key components are in place and connected in the right way!'' (P07)
\end{quote}
\paragraph{\textbf{Heuristic \#4. It is reasonable to trust agents when dealing with new information or unfamiliar contexts.}}
Participants showed signs of automation bias, trusting agents' approaches and outputs when working in new domains or dealing with new information. P14, for instance, explicitly called this out: \textit{``sometimes there are very technical details, like using certain libraries a certain way and I have no knowledge about the library. That is the hard part. I usually will have to trust the model. I never used it, so yes you are right!''} This epistemic deference suggests trust by necessity; lacking expertise and knowledge, developers move away from explicit verification to implicit reliance. Such deference may occur occasionally during a software project, but can also extend to the entire project itself:
\begin{quote}
    ``I shipped a feature in Go. I did not know Go a month ago. I'm not a fan of Go. Yeah, it worked. [...] I'm not a judge of Go, but it works. I reviewed every line and they looked legit.'' (P15)
\end{quote}
We also observed an interesting variant of `social proof' in the context of this heuristic.
When working in uncharted waters, agreement between two \textit{different} agents on the same task increases users' trust in agents:
\begin{quote}
    ``I usually use the same prompt on both [agents] just to have some more safety. If both are showing similar things, I have higher confidence it's giving me a grounded and comprehensive report.'' (P07)
\end{quote}

\section{Discussion and Implications}
\label{sec:discussion}
Our findings provide an early look into the emerging work, challenges, and heuristics of developers overseeing software agents.
Our qualitative account of the lived reality of overseeing software agents is a contribution in and of itself, providing much-needed empirical anchors to start bridging the theory and practice of agent oversight in at least one specific domain.
Below we describe two key takeaways, future research directions, and  implications for the human-centered design of software agents and for the software engineering practice.

\subsection{Oversight entails not only reviewing and monitoring, but also controlling and steering}
Our findings reveal that oversight work is likely not only reactive and evaluative, but also preventative and proactive.
\textit{Post hoc review} is the most visible form of oversight work in our study and in research \cite{spotting-mistakes-algo-2023, spotting-mistake-signal-2024, review-burden-llm-2025, passi2025_agenticoversightproblem}.
Additionally, although our participants did little \textit{real-time monitoring}, it is seen as a growing challenge for agent oversight \cite{passi2025_agenticoversightproblem, chan2024_visibilityintoagents, verhagen2025_moralagent, henzinger2023_monitoringfairness}.
However, participants described that oversight involves more than just reviewing agent outputs and monitoring agent actions.
Developers also \textit{control} and \textit{steer} agents.
Oversight work begins—as \textit{a priori control}—even before prompting.
For instance, developers anticipate failures, preemptively configuring agents to prevent them from making specific mistakes.
Oversight work is also underway—as \textit{co-planning}—before agents start executing tasks.
For instance, developers revise plans with agents and seed solutions to them, pushing agents in desirable directions away from pitfalls.

Some of the oversight work we found fits traditional notions of oversight (e.g., reviewing and monitoring), while some (e.g., preemptive configuration) appear less like oversight and more like ``good'' software engineering practices.
However, participants described doing such anticipatory ``pre'' work at times with \textit{explicit} oversight intentions (e.g., preventing known mistakes or failures from occurring altogether).
This is not to say that our findings warrant redefining oversight to now include software engineering practices.
Instead, surfacing the oversight aspirations underpinning control and steering work, we highlight that oversight work is emerging not just in expected spaces, but also elsewhere in prompts and settings.

We show different oversight work occurring \textit{throughout} human-agent interaction, expanding research scholarship on \textit{what} users do as humans-in/on/over-loops (e.g., we show that users anticipate, and not just react) and \textit{where} oversight loops begin/end (e.g., we show that they begin long before prompting) \cite{cranor2008_humanintheloop, green2019_limitsofalgodecisionmaking, cheruvu2025_inonloop, verhagen2025_moralagent, zuger2025_loopcreditlending}.
Our insights into \textit{post hoc review} and \textit{real-time monitoring} extend existing research on transparency-driven approaches to human-AI and human-agent collaboration \cite{sterz2024_interdisciplinaryhumanoversight, henzinger2023_monitoringfairness, sanneman2024_xaiinformationbottleneck}.
Our insights into \textit{a priori control} and \textit{co-planning} surface new research sites such as the role of ``anticipation'' \cite{clarke-anticipation-2016, steinhardt-anticipation, ehsan2024_seamfulxai}, the affordances and limits of natural language-based human-agent communication \cite{passi2025_agenticoversightproblem, challenges-agent-communication-2024, elizabeth-CoT-thought-2025, apple-illusion-of-thinking-2025}, and the ``managerial'' work users must do when using agents (and its impact on craftsmanship, learning, and professional identity) \cite{daly2025_managementskillagents, weidmann2025_managementleadership, tssamados2025_supervisiontoteaming}.

\subsection{Oversight functions not through exhaustive awareness, but through practical heuristics}
As explained in subsection \ref{sec:oversightofAI}, existing research largely aspires for ``effective'' oversight by facilitating continuous, ``meaningful'' human control over AI and agents \cite{filippo2018_meaningfulhumancontrol, cranor2008_humanintheloop, shneiderman2020_reliablesafeai, cavalcante2023_meaningfulhumancontrol, openai2023practices}.
However, our findings show that such aspirations often prove unscalable and are decoupled from the practical realities of users running software agents.
We found that during \textit{post hoc review} participants opted for good-enough supervision driven by ``bounded rationality'' \cite{simon1972_boundedrationality} rather than exhaustive awareness of agentic outputs and workings.
Faced with complex agentic systems, developers adopt emergent heuristics that prioritize functional efficiency over idealized perfection.
Plans become proxies for agentic working, test results stand in for code correctness, and skimming becomes the main information processing mechanism.
These are not lazy shortcuts but necessary adaptations users do to ``practically accomplish'' \cite{garfinkel1987_practicalaccomplishment, schutz1967_phenomenologypracticalaccomplishment} oversight by working \textit{around} constraints.
Agentic systems will always be ``too fast, too complex, and too much for users,'' \cite[p. 18]{passi2025_agenticoversightproblem} requiring them to inevitably use heuristics that ``satisfice'' rather than ``optimize'' oversight \cite{simon1956_satisficing, hollnagel2009_humanfactorssatisficing, pirollicard1999_informationforaging, kling1980_heuristicscommonsensen}.
We outline emergent heuristics to start reconciling theoretical work on human oversight with its situated reality \cite{chan2024_visibilityintoagents, sterz2024_interdisciplinaryhumanoversight, koulu2020_humancontroloversight}, inviting normative discussions around which heuristics are worth embracing and which introduce risks that must be mitigated \cite{green2022_oversightisflawed, tverskykahneman1974_judgmentunderuncertainty, shneiderman2022_humancenteredai}.

\subsection{Implications for the human-centered design of software agents}
While not all oversight challenges described in subsection \ref{sec:oversightforms} are immediately solvable (e.g., most need further research or technological advancements), it is possible to address certain challenges and facilitate agent oversight.

\begin{enumerate}
    \item Developers struggle during \textit{a priori control} because of insufficient information on agents and under-reliance on custom instructions. Agent builders can address this by providing configuration guides that explain how users can control specific elements of agent behavior, by surfacing constraints and defaults early to help developers anticipate downstream agent behavior, and by providing visible, low-effort ways for users to leverage custom instructions.

    \item Developers struggle during \textit{co-planning} because of difficulties formulating user goals as agent prompts and decomposing tasks optimally. Agent builders can address this by offering prompting help to users through context-aware suggestions and by providing suggested sub-tasks recommendations combined with their resource costs, reasoning depth, and execution previews.

    \item Developers struggle during \textit{real-time monitoring} to fix agent mistakes, using ad hoc cues to identify them. Agent builders can address this by designing interfaces that help users better understand and guide agent actions \cite{xie2024waitgpt} (e.g., via reasoning panels to inspect incorrect learnings and inline controls to fix them) and by surfacing contextual signals to aid issue identification (e.g., via visualizing agent memory and time data).

    \item Developers struggle during \textit{post hoc review} because of the large cognitive distance between developers and agent-generated code. Agent builders can address this by reducing friction in verifying agent outputs (e.g., augment code diff tools with intent, reasoning, and proposed tests) and by providing ways to help developers reverse engineer agent logic (e.g., interweave rationale and reasoning traces, and visualize impact of changes on agent output).
\end{enumerate}

\subsection{Implications for the software engineering practice: Individuals and organizations}The traditional `craftsman' model of software engineering is giving way to a new role: the `developer-manager', wherein hands-on coding is increasingly becoming secondary to the work of oversight.
This shift carries significant implications. 
\textit{First, for the individual developer}, the core competency is shifting from syntax mastery to architectural design, critique, and review.
This requires developers to have the ability to understand system limitations and detect anomalies (mandated by EU AI act's Article 14(4)(a) \cite{Oversight-EU-ACT}) and its success hinges on balancing managerial judgment and the technical intuition required for final vetting.
How developers take on new agent oversight work and develop new oversight skills while meaningfully preserving their coding skills will shape their professional identity \cite{futureofworkers2026}.
\textit{Second, for engineering organizations}, human oversight is evolving from a best practice into a strict legal mandate \cite{Oversight-EU-ACT}, likely creating tensions with the efficiency-driven heuristics we saw in our study.
To remain compliant, organizations must ensure developers, especially those working with high risk AI systems, possess the situational awareness and epistemic access necessary to resist automation bias in fast-paced workflows.

\section{Limitations and Future Work}
\label{sec:limitations}
This research takes important first steps and provides valuable contributions on agent oversight in practice but is not without limitations.
\textit{First,} while we tried recruiting participants from diverse professional settings, most participants worked in well-resourced technology firms (e.g., generous token limits, rich tool integrations, and early access to new features).
Monetary and computational costs mattered little in their oversight decisions.
Future work must examine oversight challenges and implications in resource-constrained settings.
\textit{Second,} 12 of our 17 participants worked at the same technology company.
Although their experiences were echoed by the 5 participants from other firms, one company's over-representation likely introduced some bias in our findings.
Future research on oversight in diverse organizational contexts can complement this research with a more holistic view of agent oversight work.
\textit{Third,} our participants delegated relatively small tasks to agents, rendering, for example, real-time monitoring optional and post hoc review easier.
Future research must analyze agent oversight work in the context of complex, longer-running tasks, while also examining how agent builders conceptualize and facilitate these more demanding forms of oversight work and requirements.

\section{Conclusion}
Our study provides an exploratory analyses of how software developers oversee coding agents in practice.
Developers perform diverse oversight work using specific practices to achieve different goals.
We identify pain points developers encounter when supervising agents and the emergent heuristics they adopt to navigate the intrinsic uncertainty of working with agents.
Our findings reveal the nuanced ways users have begun exercising oversight, beginning as early as during configuration and planning, and continuing through ongoing verification.
\textit{Doing} oversight is less of a single act and more of a coordinated set of ad-hoc evolving practices.
By starting to make these activities visible, we contribute early empirical anchors for understanding, researching, and improving agent oversight.
Our work also reflects on broader implications for tooling, individuals, organizations and future research to study, design, and support oversight in practice to enable effective collaboration and complementarity between humans and agents.

\begin{acks}
We thank our research participants for sharing their experiences with us. 
We express our gratitude to Jenna Butler for her insights during the study’s design phase, to Dean Carignan, Ben Zorn and Siddhi Tadpatrikar for their assistance with participant recruitment. 
We also thank our colleagues Amy K. Heger and Kathleen Walker for help with this research.
Last but not the least, we thank the anonymous reviewers for their constructive feedback.
\end{acks}

\section*{Generative AI Usage}
The authors used generative AI tools for transcribing interviews, proofreading, and at times to improve the fluency of sentences. Generative AI was not used for data analysis or generating technical content.

\bibliographystyle{ACM-Reference-Format}
\bibliography{sample-base}
\newpage
\appendix
\section{Appendix: Participant demographics} 
\label{sec:appendix_participant_demograhics} 
We provide a detailed table below that captures all participants' demographics, including their self-reported programming experience, software agent experience, usage.
\label{sec:appendix_participant_demograhics} 
\setlength{\tabcolsep}{1.5pt}
\small
{
\begin{longtable}{cccl@{\hspace{4pt}}l@{\hspace{5pt}}l@{\hspace{5pt}}l@{\hspace{8pt}}l}
\hline
\textbf{No.} &
  \textbf{Gender} &
  \textbf{Country} &
  \textbf{Role} &
  \textbf{\begin{tabular}[c]{@{}l@{}}Programming \\ experience\end{tabular}} &
  \textbf{\begin{tabular}[c]{@{}l@{}}Software agent \\ experience\end{tabular}} &
  \textbf{\begin{tabular}[c]{@{}l@{}}Software agent \\ usage frequency\end{tabular}} &
  \textbf{\begin{tabular}[c]{@{}l@{}}Software agents \\ used\end{tabular}} \\ \hline
\endfirsthead
\endhead
\hline
\endfoot
\endlastfoot
P01 &
  F &
  US &
  \begin{tabular}[c]{@{}l@{}}Software engineer / \\ developer\end{tabular} &
  \textgreater 10 years &
  Experienced &
  Once a week &
  GitHub Copilot \\
P02 &
  M &
  US &
  \begin{tabular}[c]{@{}l@{}}Software engineer/\\ developer\end{tabular} &
  7-10 years &
  \begin{tabular}[c]{@{}l@{}}Very \\ experienced\end{tabular} &
  Daily &
  \begin{tabular}[c]{@{}l@{}}Cline; Cursor; \\ GitHub Copilot\end{tabular} \\
P03 &
  M &
  US &
  Engineering manager &
  \textgreater 10 years &
  Experienced &
  Daily &
  \begin{tabular}[c]{@{}l@{}}Claude Code; Cline; \\ Cursor; GitHub Copilot\end{tabular} \\
P04 &
  M &
  India &
  \begin{tabular}[c]{@{}l@{}}Software engineer/\\ developer\end{tabular} &
  1-3 years &
  \begin{tabular}[c]{@{}l@{}}Very \\ experienced\end{tabular} &
  Daily &
  \begin{tabular}[c]{@{}l@{}}Cursor; GitHub Copilot; \\ OpenAI Codex\end{tabular} \\
P05 &
  M &
  US &
  \begin{tabular}[c]{@{}l@{}}Technical Program \\ Manager\end{tabular} &
  1-3 years &
  Experienced &
  \begin{tabular}[c]{@{}l@{}}Several times \\ a week\end{tabular} &
  \begin{tabular}[c]{@{}l@{}}Amazon Kiro; Cursor; \\ GitHub Copilot\end{tabular} \\
P06 &
  M &
  US &
  \begin{tabular}[c]{@{}l@{}}Data science/ \\ ML engineering\end{tabular} &
  4-6 years &
  \begin{tabular}[c]{@{}l@{}}Somewhat \\ experienced\end{tabular} &
  \begin{tabular}[c]{@{}l@{}}Several times \\ a week\end{tabular} &
  Cursor \\
P07 &
  M &
  US &
  Engineering manager &
  \textgreater 10 years &
  Experienced &
  Daily &
  \begin{tabular}[c]{@{}l@{}}Cline; GitHub Copilot; \\ Rubber Ducky\end{tabular} \\
P08 &
  M &
  US &
  \begin{tabular}[c]{@{}l@{}}Software engineer/\\ developer\end{tabular} &
  \textgreater 10 years &
  Experienced &
  Daily &
  GitHub Copilot \\
P09 &
  M &
  Brazil &
  \begin{tabular}[c]{@{}l@{}}Software engineer/\\ developer\end{tabular} &
  7-10 years &
  \begin{tabular}[c]{@{}l@{}}Very \\ experienced\end{tabular} &
  Daily &
  \begin{tabular}[c]{@{}l@{}}Amazon CodeWhisperer; \\ GitHub Copilot; \\ OpenAI Codex\end{tabular} \\
P10 &
  M &
  US &
  \begin{tabular}[c]{@{}l@{}}Software engineer/\\ developer\end{tabular} &
  4-6 years &
  Experienced &
  \begin{tabular}[c]{@{}l@{}}Several times \\ a week\end{tabular} &
  Gemini; GitHub Copilot \\
P11 &
  M &
  US &
  \begin{tabular}[c]{@{}l@{}}Software engineer/\\ developer\end{tabular} &
  \textgreater 10 years &
  \begin{tabular}[c]{@{}l@{}}Somewhat \\ experienced\end{tabular} &
  Daily &
  \begin{tabular}[c]{@{}l@{}}Amazon CodeWhisperer;  \\ Cursor; GitHub Copilot\end{tabular} \\
P12 &
  M &
  US &
  \begin{tabular}[c]{@{}l@{}}Software engineer/\\ developer\end{tabular} &
  \textgreater 10 years &
  \begin{tabular}[c]{@{}l@{}}Somewhat \\ experienced\end{tabular} &
  \begin{tabular}[c]{@{}l@{}}Several times \\ a week\end{tabular} &
  Gemini \\
P13 &
  M &
  US &
  \begin{tabular}[c]{@{}l@{}}Software engineer/\\ developer\end{tabular} &
  \textgreater 10 years &
  Experienced &
  Daily &
  Cursor; GitHub Copilot \\
P14 &
  F &
  US &
  \begin{tabular}[c]{@{}l@{}}Software engineer/\\ developer\end{tabular} &
  \textgreater 10 years &
  Experienced &
  Daily &
  \begin{tabular}[c]{@{}l@{}}Aider; Claude Code;\\ GitHub Copilot\end{tabular} \\
P15 &
  M &
  US &
  \begin{tabular}[c]{@{}l@{}}Software engineer/\\ developer\end{tabular} &
  \textgreater 10 years &
  Experienced &
  Daily &
  \begin{tabular}[c]{@{}l@{}}Claude Code, \\ GitHub Copilot; VsCode;\end{tabular} \\
P16 &
  M &
  US &
  \begin{tabular}[c]{@{}l@{}}Software engineer/\\ developer\end{tabular} &
  \textgreater 10 years &
  \begin{tabular}[c]{@{}l@{}}Very \\ experienced\end{tabular} &
  Daily &
  \begin{tabular}[c]{@{}l@{}}Codex CLI; Cursor; \\ GitHub Copilot\end{tabular} \\
P17 &
  M &
  US &
  \begin{tabular}[c]{@{}l@{}}Software engineer/\\ developer\end{tabular} &
  \textgreater 10 years &
  \begin{tabular}[c]{@{}l@{}}Somewhat\\ experienced\end{tabular} &
  Daily &
  \begin{tabular}[c]{@{}l@{}}Claude Code; Cursor; \\ GitHub Copilot\end{tabular} \\ \hline
\\
\caption{Participant details (self-reported): Demographics, programming experience, software agent experience, software agent usage frequency, and software agents used.}
\label{tab:my-table}\\
\end{longtable}
}
\normalcolor

\section{Appendix: Agent usage scenarios}
\label{sec:appendix_agent_usage_scenarios}
\begin{figure}[h!]
\centering
 \includegraphics[scale=0.5]{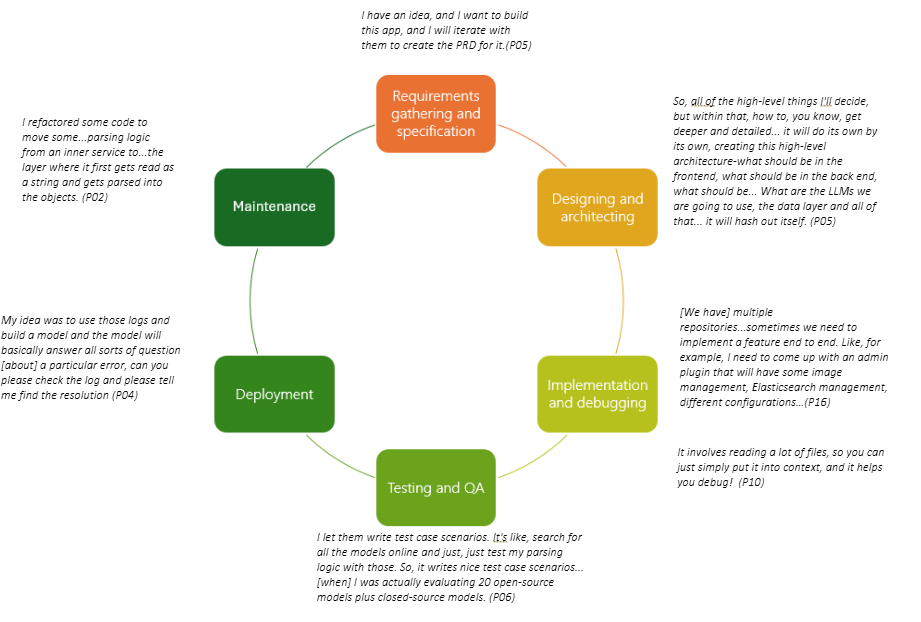}
  \caption{Participants' description of scenarios in which they use AI agents for software development}
  \Description{Quotes for agent usage}
  \label{fig:sdlc}
\end{figure}

Our study did not aim to quantify software agent usage patterns, but participants shared their real-world situations to illustrate and ground their oversight practices. We illustrate key examples in Figure \ref{fig:sdlc}. Participants' accounts revealed that AI agents are integrated across several core phases of developers' software development and engineering practices, automating various tasks such as drafting and validating product requirements documents, implementing end-to-end features, debugging code, analyzing storage logs, and code refactoring. Participants used agents across codebases implemented in different programming languages, including Python, Java, Go, and C\#. These descriptions complement industry reports \cite{anthropicAnthropicEconomic, perplexityPeopleAgents} demonstrating agents pervasive use in automating a variety of programming tasks.

Software development is not limited to writing code or implementing features; it also involves continuously understanding and learning code \cite{xia2018measuringprogramcomprehension, dualaekoko2012askingquestions} and collaborating with other developers \cite{latoza2006maintainingmentalomdels}.
Our participants also used agents to support these activities.
For instance, P16 used agents to catch-up on codebase changes after returning from paternity leave, while P06 leveraged agents to generate visualizations to communicate trends in their research findings. 

One interesting insight from our data is that when working with production-grade enterprise software, participants usually delegated only small, well-bounded tasks to software agents--tasks that an agent can complete in a short execution window typically within a ``few dozen seconds to a minute'' (P12). This has implications for developers’ oversight practices. In a few cases, agent run time was relatively longer: for proof-of-concept or prototype work (e.g., 20-25 minutes (P10)) and when participants were advanced power users (e.g., 15-20 minutes (P16)) who engaged agents in more extended tasks.

\section{Appendix: Interview Topic Guide}
\label{sec:appendix_interview_topic_guide}

\subsubsection{Introduction and Welcome}
Thank you for participating in this study. We are interviewing people to understand how they use and oversee agents. Your participation is voluntary, and your responses will be anonymous and confidential. Results will be reported only in aggregate and when referred, your quotes will be de-identified. With your permission, Marvin will be recording this meeting to ensure accuracy. 

\subsubsection{Agent usage}
\begin{enumerate}
    \item To begin, could you describe the scenarios in which you use coding agents? e.g., Github Copilot (agent mode), Cursor, Claude Code, Windsurf, OpenAI Codex, Mistrals Devstral (Open source)
\end{enumerate}

\subsubsection{Before deployment}
\begin{enumerate}
    \item How do you decide which agent(s) to choose for which task i.e. which factors help you decide the agent you work with?
    \begin{itemize}
        \item Do you look at leaderboards, model cards, or any benchmarks e.g., SWE-bench when selecting models to use, their training cut off?
    \end{itemize}
    \item Are there any tradeoffs that you encounter when choosing agents?
    \item How have you learned to tailor your agent for your software project?
     \begin{itemize}
        \item e.g., Custom instructions – to include best practices for your applications stack?
          \item e.g., Connections to things like databases with MCP? Which MCP server to use, the tools available from it?
     \end{itemize}
    \item What risks related to agentic automation do you consider during this initial phase?
    \item What evals do you put in place? What goals do they serve?
    \begin{itemize}
        \item agents’ accuracy – consistency with initial requirements?
    \item agents’ biases in reasoning or execution
    \item agents’ time to execution
    \item agents’ adherence to intended scope
    \item agent not going off-task
    \item agent not leaking sensitive information
    \item agent not producing copyright information
    \end{itemize}
    
\end{enumerate}
\subsubsection{At run-time}
\begin{enumerate}
    \item What do you do when the agents are deployed?
    \item What risks related to agentic automation do you consider when the agent is executing on a task?
    \item How do you govern your agent(s) when they are deployed?
    \item How do you intervene agents? Could you describe the types of interventions you make?
\end{enumerate}

\subsubsection{Post deployment}
\begin{enumerate}
    \item How do you test the correctness of the agents’ actions and outputs?
    \item How do you define goodness for agents’ actions and outputs?
    \item Which artifacts do you inspect to determine this goodness?
    \item What verification tools are available to you? e.g., manual, automated techniques, agents-as-judge, confidence scores, agents’ justifications, etc.
\end{enumerate}

\subsubsection{Agent impact on workflow}
\begin{enumerate}
    \item Do you ever feel stuck when using agents? What things do you do to get unstuck? Are you own our own, with your rubber duck, or with another team member when you troubleshoot?
    \item How has your workflow changed, now that you have access to coding agents?
    \begin{itemize}
        \item What activities do you do now that you didn’t do earlier?
        \item What activities do you not do when you are working with agents that you would otherwise do?
    \end{itemize}    
\end{enumerate}

\subsubsection{Concluding remarks}
I'd like to open it up for questions that you may have. You will receive the \$40 giftcard in the email you provided earlier. Thank you for your time.

\section{Appendix: Four forms of oversight work}
\label{sec:appendix_forms_of_oversight_table}
Table \ref{tab:Table Four Forms of Oversight} captures the four forms of emergent oversight work and their descriptions as revealed in our exploratory analysis. 
\begin{table}[H]
\centering
\setlength{\tabcolsep}{12pt}
{

\begin{tabular} {p{0.15\textwidth}|p{0.75\textwidth}}
 \toprule
 \multicolumn{1}{c}{\textbf{Oversight Work}} & \multicolumn{1}{c}{\textbf{Description}} \\
 \toprule

\textbf{A priori}\newline {\textbf{control}} & A preventative form of oversight work that involves giving instructions to agents to direct and limit their workings \textit{before} delegating tasks to them. A priori control includes configuring agent settings, constraining agent boundaries, and supplying global context.\\
\midrule

\textbf{Co-planning}\newline {\textbf{work}} & A proactive form of oversight work that involves pre-execution goal-setting activities that agents and developers jointly perform to establish common ground. Examples include drafting plans with agents, iterative prompting, and seeding partial solutions to agents.\\
\midrule

\textbf{Real-time}\newline {\textbf{monitoring}} & A reactive form of oversight work that involves observing agents and intervening, when needed, as they perform tasks. Examples includes observing agent reasoning and pausing/stopping agents mid-execution.\\
\midrule

\textbf{Post-hoc}\newline {\textbf{review}} & An evaluative form of oversight work that involves reviewing and correcting agent outputs after execution. Examples include verifying outputs, making adjustments when agents falter, and re-prompting to fix mistakes.\\
\bottomrule
\end{tabular}
 }
\caption{The four forms of emergent agent oversight work}
\label{tab:Table Four Forms of Oversight}
\end{table}
\normalcolor
\end{document}